\documentclass[aps,prl,reprint,superscriptaddress,floatfix,longbibliography]{revtex4-1}

\usepackage{bm}
\usepackage{amsmath}
\usepackage{graphicx}
\usepackage{siunitx}
\usepackage{url}
\usepackage{hyperref}
\usepackage{etoolbox}
\apptocmd{\sloppy}{\hbadness 10000\relax}{}{}
\graphicspath{{figures/}}
\begin{document}

\title{Measuring environmental quantum noise exhibiting a non-monotonous
spectral shape}
\author{Y. Romach\normalfont\textsuperscript{\dag}}
\email[Corresponding author: ]{yoav.romach@mail.huji.ac.il}
\affiliation{The Racah Institute of Physics, The Hebrew University of Jerusalem, Jerusalem 9190401, Israel}
\author{A. Lazariev}
\thanks{These authors contributed equally to this work.}
\affiliation{Department of NanoBiophotonics, Max Planck Institute for Biophysical Chemistry, Am Fassberg 11, 37077 G{\"o}ttingen, Germany}
\author{I. Avrahami}
\affiliation{Department of Applied Physics, Rachel and Selim School of Engineering, The Hebrew University of Jerusalem, Jerusalem 9190401, Israel}
\author{F. Klei{\ss}ler}
\affiliation{Department of NanoBiophotonics, Max Planck Institute for Biophysical Chemistry, Am Fassberg 11, 37077 G{\"o}ttingen, Germany}
\author{S. Arroyo-Camejo}
\affiliation{Department of NanoBiophotonics, Max Planck Institute for Biophysical Chemistry, Am Fassberg 11, 37077 G{\"o}ttingen, Germany}
\author{N. Bar-Gill}
\affiliation{The Racah Institute of Physics, The Hebrew University of Jerusalem, Jerusalem 9190401, Israel}
\affiliation{Department of Applied Physics, Rachel and Selim School of Engineering, The Hebrew University of Jerusalem, Jerusalem 9190401, Israel}

% Expressions
\newcommand{\tauCPMG}{\tau_{\textrm{free}}}
\newcommand{\tauFREE}{\tau_{\textrm{free}}}
\newcommand{\tCPMG}{t_{\textrm{CPMG}}}
\newcommand{\tDYSCO}{t_{\textrm{DYSCO}}}
\newcommand{\FWHM}[1]{\textrm{FWHM}_{\textrm{#1}}}
\newcommand{\fMax}[1]{f_{\textrm{max}}^{\textrm{#1}}}
\newcommand{\fMin}[1]{f_{\textrm{min}}^{\textrm{#1}}}
\newcommand{\fRabi}{f_{\textrm{Rabi}}}

% Show all comments
\newcommand{\strike}[1]{{\color{red}\sout{#1}}}
\newcommand{\place}[1]{{\color{red}#1}}
\newcommand{\FK}[1]{{\color{blue}#1}}
\newcommand{\YR}[1]{{\color{green}#1}}
\newcommand{\comment}[1]{{\color{cyan}#1}}
% Produce plain text
%\newcommand{\strike}[1]{}
%\newcommand{\place}[1]{{\color{black}#1}}
%\newcommand{\FK}[1]{}
%\newcommand{\YR}[1]{}
%\newcommand{\comment}[1]{}

\begin{abstract}
Understanding the physical origin of noise affecting quantum systems is important for nearly every quantum application. Quantum noise spectroscopy has been employed in various quantum systems, such as superconducting qubits, NV centers and trapped ions. Traditional spectroscopy methods are usually efficient in measuring noise spectra with mostly monotonically decaying contributions. However, there are important scenarios in which the noise spectrum is broadband and non-monotonous, thus posing a challenge to existing noise spectroscopy schemes.
Here, we compare several methods for noise spectroscopy: spectral decomposition based on the Carr-Purcell-Meiboom-Gill (CPMG) sequence, the recently presented DYnamic Sensitivity COntrol (DYSCO) sequence and a modified DYSCO sequence with a Gaussian envelope (gDYSCO).
The performance of the sequences is quantified by analytic and numeric determination of the frequency resolution, bandwidth and sensitivity, revealing a supremacy of gDYSCO to reconstruct non-trivial features.
Utilizing an ensemble of nitrogen-vacancy centers in diamond coupled to a high density $^{13}$C nuclear spin environment, we experimentally confirm our findings.
The combination of the presented schemes offers potential to record high quality noise spectra as a prerequisite to generate quantum systems unlimited by their spin-bath environment.

\end{abstract}

\maketitle
\section{Introduction}

Quantum systems are inherently subject to noise originating from their coupling to the environment, which in turn affects their coherence properties, with implications to quantum information processing, many-body dynamics and quantum sensing \cite{laddquantum2010,Shnirman_TLS_decoherence,Ithier_Decoherence,Villar_Decoherence,Martinis_noise1,Martinis_noise2}. As a consequence, studying the noise sources affecting quantum systems and optimizing schemes for mitigating them has been of great interest over the past decade \cite{klauder_anderson,marcus_dd_scaling,Martinis_noise1,Martinis_noise2,yan_rotating-frame_2013,Ithier_Decoherence,Villar_Decoherence,oliver,romach2015}. Moreover, devising new quantum noise spectroscopy methods (general techniques for analysing noise sources using quantum probes) is a fundamental aspect of the field and has attracted significant attention over the past years\cite{romach2015,bar-gill_spect_decomp,oliver, yan_rotating-frame_2013,farfurnik_experimental_2017}.

Quantum noise spectroscopy has advanced in recent years, and has been employed in the context of various quantum systems, such as superconducting qubits \cite{oliver}, trapped ions \cite{Kotler2011}, $^{13}$C atoms in adamantane \cite{alvarez_measuring_2011}, optically trapped ultra-cold atoms \cite{almog_direct_2011} and nitrogen-vacancy (NV) centers in diamond \cite{bar-gill_spect_decomp,romach2015}. These studies have led to a deeper understanding of the physical origins and dynamics of the noise sources and of the system-environment interaction (e.g. \cite{bar-gill_spect_decomp,romach2015,oliver}), as well as to advanced sensing applications (e.g. \cite{Kotler2011}).

In most of the previous works, the spectrum of the relevant environmental noise was either nearly monotonically decreasing (e.g. $1/f$ noise \cite{oliver}, DC-centerd Lorentzian \cite{bar-gill_spect_decomp,romach2015}), bandwidth-limiting techniques were employed \cite{bar-gill_spect_decomp} or the noise spectroscopy was performed using straightforward continuous driving, resulting, again, in limited bandwidth (e.g. \cite{almog_direct_2011}). These approaches are therefore limited in addressing important situations commonly encountered in realistic systems, in which the contributing noise is distributed over a large frequency bandwidth but the noise spectrum also exhibits strong resonant features (and thus a non-monotonous spectrum).

Here, we investigate the potential of noise spectroscopy based on three microwave (MW) driving sequences: the Carr-Purcell-Meiboom-Gill (CPMG) sequence post-processed by spectral decomposition (CPMG SD)\cite{bar-gill_spect_decomp,romach2015}, and the two recently introduced sequences: DYnamic Sensitivity COntrol (DYSCO) and a modified DYSCO scheme with a Gaussian envelope (gDYSCO)\cite{andrii_DYSCO}. 
The properties of the sequences are studied analytically and numerically in terms of accessible bandwidth, frequency resolution and gain as well as their implications for the reconstruction of noise spectra.
Experiments were conducted utilizing an ensemble of NV centers in diamond coupled to a bath of $^{13}$C nuclear spins in order confirm the predicted behavior.

The NV center is a defect in the diamond lattice, in which one of the carbon atoms is replaced with a nitrogen atom and an adjacent site is substituted by a vacancy. A zero-field splitting of $\sim \SI{2.87}{\GHz}$ between $m_s = 0$ and $m_s = \pm 1$ defines the triplet ground state. By applying a static magnetic field one can break the degeneracy between the $m_s = \pm1$ sub-states through the Zeeman effect, thus creating an effective two-level system. The NV center electron spin can be initialized and detected optically due to spin-dependent transitions, and coherently manipulated within the ground-state spin manifold utilizing MW fields \cite{Doherty2013}.

In the last decade, the NV color center in diamond has emerged as a leading platform for quantum information, quantum meteorology and magnetic sensing \cite{laddquantum2010,taylor2008,walsworth_imager,wrachtrup_efield,Mamin2013,Staudacher2013,Grinolds2013,rondin13} due to its remarkable properties such as long coherence times at room temperature \cite{Doherty2013, Bar-Gill2013}. Methods adopted from the field of nuclear magnetic resonance, such as dynamical decoupling \cite{hanson2010,naydenov,suterKDD,cory}, increased the coherence time even further \cite{Bar-Gill2013}.
Substantial research efforts are invested in order to understand the relevant noise sources affecting NV centers and their physical origins, as well as to optimize protocols to suppress their adverse effects\cite{romach2015,kim_decoherence_2015,alvarez_measuring_2011}.

Recently, NV centers have been used to measure sub-millihertz NMR spectra \cite{schmitt_submillihertz_2017, boss_quantum_2017, glenn_high-resolution_2018}. In these works, the authors took advantage of the long coherence times of nuclear spins, and/or implemented a locked-in or quantum-homodyne method in order to achieve enhanced spectral resolution. In \cite{glenn_high-resolution_2018}, the authors also applied a driving pulse on the nuclear spins in order to further increase the resolution. However, these measurements result in a relatively low spectral bandwidth, which is insufficient for a general noise spectroscopy method, aiming to reconstruct the entire noise spectrum, as is the focus here.

The system's decoherence due to environmentally-induced noise can be expressed as $C(t)=e^{-\chi(t)}$, where $\chi(t)$ contains the dependence of the decoherence processes on the spectral noise density $S(\omega)$ through \cite{dassarma_dd}:
\begin{equation}
\chi(t)=\int_{0}^{\infty}\frac{d\omega}{\pi}S(\omega)\frac{\tilde{F}(\omega t)}{\omega^2}
\label{eq:chi-pre}
\end{equation}
Here, $F(\omega t)$ is the sequence dependent frequency filter function (FF) defined as the absolute square of the Fourier transform of the time-dependent sequence sensitivity function.
For convenience, we define $\textrm{FF}(\omega t) = \frac{2 F(\omega t)}{t \pi \omega^2}$ such that $\int_{0}^{\infty} \mathrm{FF}(\omega t) \mathrm{d}\omega = 1$ (for a CPMG experiment). Eq.\,(\ref{eq:chi-pre}) becomes:
\begin{equation}
\chi(t)=\frac{t}{2}\int_{0}^{\infty} d\omega S(\omega) \textrm{FF}(\omega t)
\label{eq:chi}
\end{equation}

Generally, the system's sensitivity can be modulated in a pulsed or continuous manner.
While pulsed sequences (e.g. CPMG) flip the sensitivity between $\pm 1$, the continuously driven DYSCO scheme enables arbitrary modulation of the sensitivity.
Precise driving of the quantum sensor enables tuning of the FF in order to measure the noise environment at well defined sensing frequencies $f_0$. The full noise spectrum $S(\omega)$ can be reconstructed out of a set of measurements \cite{dassarma_dd}.

\begin{figure}[htbp]
	\begin{center}
 		\includegraphics[width=1 \columnwidth]{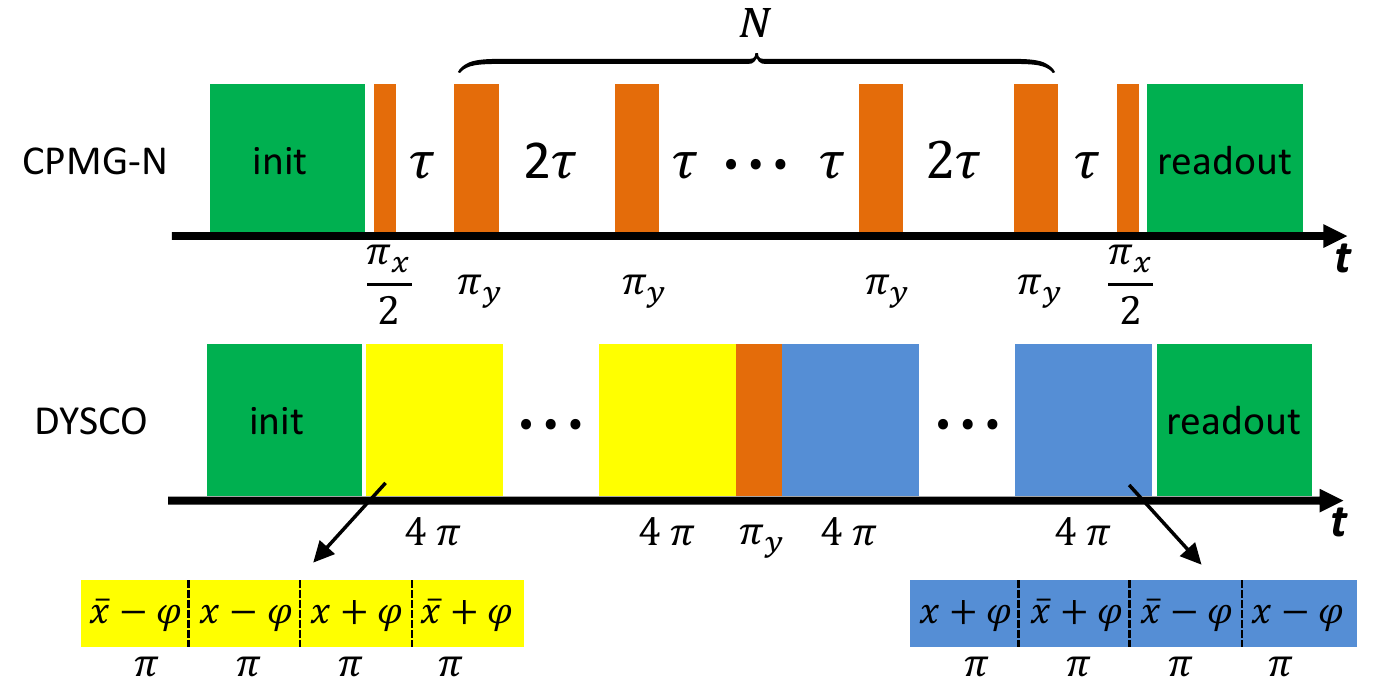}
		\protect\caption{
		Schematic of the CPMG and DYSCO control schemes: green indicates a laser pulse, orange indicates a $\pi$ or $\frac{\pi}{2}$ pulse, while yellow and blue indicate MW pulse blocks composed of $4\pi$ pulses with the denoted phases (in which $\varphi$ is a function of the block number).
		} \label{fig:fig1}
	\end{center}
\end{figure}

\section{Filter function analysis}
One of the simplest pulsed sensitivity modulation sequences is the Hahn-Echo sequence \cite{hahn}, in which a single $\pi$ pulse is applied in the middle of the sequence of length $t_{\textrm{Hahn}} = 2 \tauFREE$ in order to refocus the NV spin and eliminate DC effects.
Additionally, any noise contribution whose correlation time is shorter than the free evolution time $\tauFREE$ cancels out.
As an extension of the Hahn-Echo experiment, the CPMG-N pulse sequence \cite{cpmg} consists out of $N$ equally spaced $\pi$ pulses with $2 \tauFREE$ time intervals between them (Fig.\,\ref{fig:fig1}).
By changing the free precession time $\tauCPMG$, the experiment time $\tCPMG$ and the number of pulses $N$, the sensing frequency can be adjusted according to:
\begin{equation}
f_0^{\textrm{CPMG}}(\tauCPMG) \approx \frac{N}{2 \tCPMG} = \frac{1}{4 \tauCPMG}
\label{eq:eq2}
\end{equation}

\begin{figure}[htbp]
	\begin{center}
 		\includegraphics[width=1 \columnwidth]{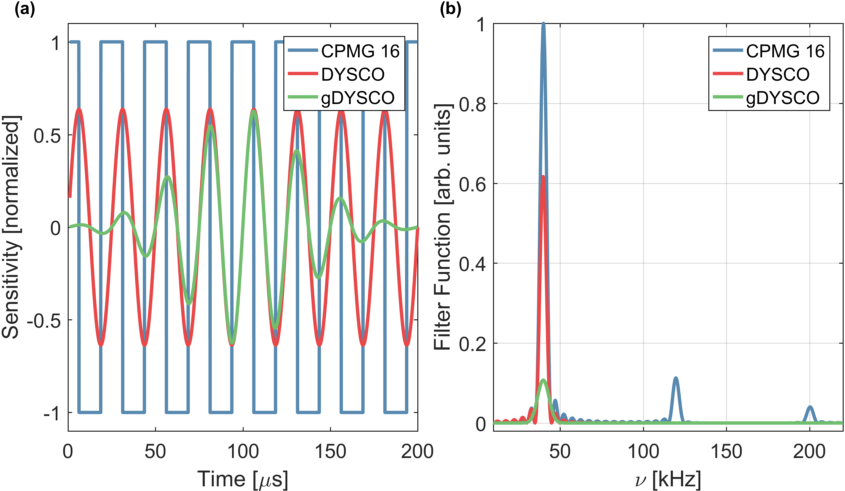}
		\protect\caption{
		The time domain sensitivity functions for CPMG-16, DYSCO and gDYSCO (a) and the respective FFs (b). Notice that the higher harmonics in CPMG disappear in DYSCO at the costs of a smaller maximal filter function height. When a Gaussian envelope is added, the side lobes disappear at the cost of an even further reduced maximal filter function height and an increased width of the main peak.
		} \label{fig:fig2}
	\end{center}
\end{figure}

The total experiment time is given by $\tCPMG = 2 N \tauCPMG$ and is ultimately limited by $2 T_1$ \cite{sousa2009}.
However, the pulsed nature of the sequence, i.e. stepwise sensitivity modulation (Fig.\,\ref{fig:fig2}(a)), introduces higher harmonics into the filter function (Fig.\,\ref{fig:fig2}(b)).
A significant contribution to the decoherence can originate from these higher harmonics if the noise spectrum is not monotonically decreasing.
While the spectral decomposition scheme incorporates precise knowledge of the FF, the reconstructed spectrum nevertheless suffers from artifacts of this origin. (See appendix for a complete description of the spectral decomposition method).

The DYSCO pulse sequence was first presented by Lazariev et al. \cite{andrii_DYSCO} as a means for selective radio-frequency (RF) spectroscopy using NV centers.
Contrary to spin-flipping sequences, DYSCO allows for control of the instantaneous sensitivity of the spin-sensor by precise pulse phase handling.
This is achieved at the cost of permanent driving and reduction of the maximal sensitivity by a factor of $\frac{2}{\pi}$.
Modulating the sensitivity function in a sinusoidal manner (Fig.\,\ref{fig:fig2}(a)) results in a filter function free of higher harmonics.
Given the finite experiment time, the DYSCO filter function is a squared sinc function that has small side lobes (Fig.\,\ref{fig:fig2}(b)). This effect can be suppressed by adding a Gaussian envelope to the sensitivity modulation (gDYSCO) (Fig.\,\ref{fig:fig2}(a)), which removes the side lobes at the cost of a further reduced sensitivity and a slightly increased width of the main peak (Fig.\,\ref{fig:fig2}(b)).

For a quantitative analysis of the FFs, we examined four parameters: resolution, minimum frequency $\fMin{}$, maximum frequency $\fMax{}$ and the gain $\Sigma$ (a measure for the system's response at the desired frequency).
While $\fMin{}$ and $\fMax{}$ define the system's bandwidth, we introduce the FWHM of the FF main peak as a quantitative measure of the frequency resolution.
However, it should be stressed that the FF side lobes (in CPMG and DYSCO) limit the resolution by increasing the effective envelope around the main peak. Therefore, the FWHM only gives a lower bound on the resolution.

The DYSCO FF is a squared sinc function such that the $\FWHM{DYSCO} \approx \frac{0.884}{\tDYSCO}$ (calculated numerically), where $\tDYSCO$ is the the experiment time, ultimately limited by $T_{1\rho}$ (the relaxation time in the driven system).
On the other hand, the gDYSCO FF is a Gaussian with $\text{FWHM}_{\text{gDYSCO}} = \frac{\sqrt{\text{ln}2}}{\pi \sigma} \approx \frac{0.265}{\sigma}$, where $\sigma$ is the width of the time-domain Gaussian envelope.
$\sigma$ is limited by the experiment time, and in our experiments and simulation it was chosen to be $\sigma = \frac{\tDYSCO}{6}$ such that the envelope is $\rightarrow 0$ at the beginning and end of the sequence.
This results in a full width half maximum of $\FWHM{gDYSCO} \approx \frac{1.59}{\tDYSCO}$.
In contrasts, the CPMG FF main peak frequency and FWHM can only be derived numerically.
The width is given by $\FWHM{CPMG} \approx \frac{0.89}{\tCPMG}$.
It is important to note that there is a significant difference between the CPMG experiment time, $\tCPMG$, and the DYSCO experiment time, $\tDYSCO$.
In DYSCO, $\tDYSCO$ remains constant regardless of the sensing frequency. However, in CPMG, $\tCPMG$  changes with the sensing frequency, which is given in Eq.\,(\ref{eq:eq2}).
It follows that the FWHM can also be written as: $\FWHM{CPMG} \approx 1.77\frac{f_0}{N}$ or $\FWHM{CPMG} \approx \frac{0.89}{2\tauCPMG N}$.
In addition, $\tDYSCO$ is limited by $T_{1\rho}$ due to its continuously driven nature while $\tCPMG$ is limited by $T_2$ (the transverse, spin-spin, relaxation time), which depends on $N$ and is ultimately limited by $T_1$ (the longitudinal, spin-lattice, relaxation time).
These non-trivial relations complicate the comparison between DYSCO and CPMG. However, it can be generally stated that CPMG can achieve a higher resolution (narrower FF) for a given frequency by increasing the number of pulses.

The gain $\Sigma$ is defined by the integral over the main peak of the FF in the region of the FWHM:
\begin{equation}
\Sigma = \int_{f_0-\Delta}^{f_0+\Delta} \textrm{FF}(f) \mathrm{d}f
\end{equation}
Where $f_0$ is the FF peak frequency and $2 \Delta$ is the FWHM. The gain, $\Sigma$, is a measure of how much the coherence curve is affected by the presence of noise around the sensed frequency and is only weakly dependent on $f_0$.
The DYSCO gain $\Sigma_{\textrm{DYSCO}}$ is approximately $60\,\%$ \cite{andrii_DYSCO} of the CPMG gain $\Sigma_{\textrm{CPMG}}$, and the gDYSCO gain $\Sigma_{\textrm{gDYSCO}}$ is approximately $20\,\%$.

The maximal frequency for CPMG is limited by the requirement $2 \tauCPMG \gg \tau_{\pi}$, where $\tau_{\pi}$ is the duration of a $\pi$-pulse. 
As a consequence, the maximal frequency is given by $\fMax{CPMG} \ll \fRabi$ (with $\fRabi$ being the Rabi/driving frequency). On the other hand, in DYSCO, as well as in gDYSCO, the maximal frequency is limited by the quantization of the sensitivity sine function (due to discrete phase steps). Quantizing the sine to $n$ steps, where $n \gg 1$, leads to $n 4 \tau_{\pi} \fMax{DYSCO} \ll 1 \Rightarrow \fMax{DYSCO} \ll \frac{\fRabi}{2 n}$.
Even though the expressions for $\fMax{}$ are similar, it needs to be stated that for small $n$ the DYSCO sensitivity modulation approaches the step like case of CPMG.
The reduced DYSCO gain, $\Sigma_{\textrm{DYSCO}}$, in combination with the appearance of higher harmonics for small $n$, reveals the supremacy of the CPMG SD method to reconstruct the noise at high frequencies.
The minimum frequency $\fMin{CPMG}$ for CPMG is estimated in a similar manner and is $\fMin{CPMG} \lesssim \frac{1}{2 T^{\text{Echo}}_2}$ (see appendix for derivation).
For DYSCO and gDYSCO the minimum frequency is given by the demand to fit at least one full period of the sensed frequency in the total experiment time so $\fMin{DYSCO} \tDYSCO > 1 \Rightarrow \fMin{DYSCO} > \frac{1}{\tDYSCO}$. Recall that the time scale in DYSCO is limited by $T_{1\rho}$.
Therefore, CPMG has a clear advantage in terms of bandwidth, or dynamic range.

\section{Simulations}

\begin{figure}[htbp]
	\begin{center}
		\includegraphics[width=1 \columnwidth]{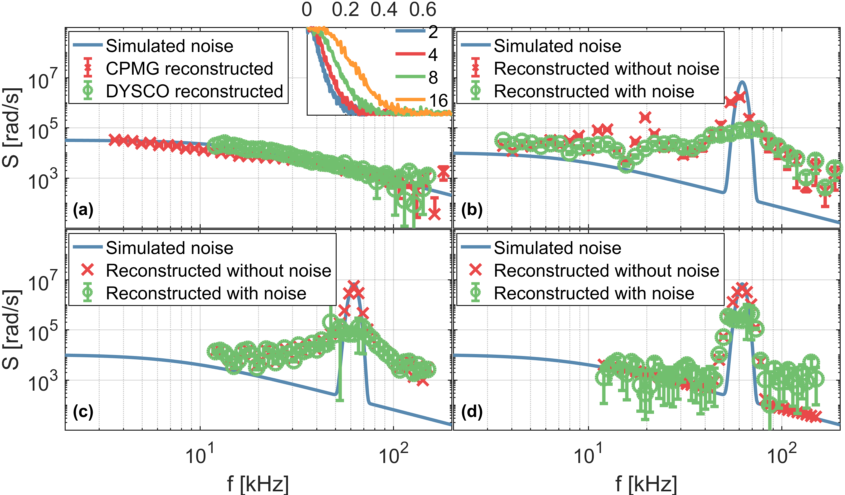}
		\caption{
		Simulations of the various reconstructions methods: (a) A noise spectrum is defined as a DC-centered Lorentzian (blue). Based on the noise spectrum, CPMG-N decoherence curves are generated (inset: numbers are N, X axis is time in ms, Y axis is coherence in arb. units). Normally distributed noise added to the coherence curves simulates experimental uncertainties. Applying the decomposition code to the coherence curves reconstructs the noise spectrum (red).
        Equivalently, the reconstruction of the noise spectrum for the DYSCO sequence is simulated (green).
        (b-d) Extends the analysis to noise spectra featuring a pronounced peak for CPMG-N, DYSCO and gDYSCO, respectively. The spectra obtained from a noise-free coherence curve are presented by the red data points whereas the data presented by the green points incorporate the simulated experimental uncertainty. In general, the consideration of experimental uncertainties blurs the noise peak.
        Higher harmonics of the CPMG-N FF cause artifacts in the reconstructed spectrum at roughly $\SI{30}{\kHz}$ and $\SI{15}{\kHz}$, which do not appear in the DYSCO and gDYSCO measurements.
        gDYSCO, which avoids the side lobes in the FF, reconstructs the noise spectrum with highest precision.
        }\label{fig:fig3}
	\end{center}
\end{figure}

Numerical simulations demonstrate the limitations of the different methods.
As a proof of principle, the trivial case of a monotonically decreasing noise spectrum $S(\omega)$ is considered.
By integrating Eq.\,(\ref{eq:chi}) for a set of CPMG FFs (see appendix for more information), complete coherence curves are simulated.
Adding normally distributed noise to the coherence curves generates realistic data sets with experimental uncertainties (inset of Fig.\,\ref{fig:fig3}(a)), which are subsequently analyzed by the CPMG SD method in order to reconstruct the spectral noise density.
The results plotted in Fig.\,\ref{fig:fig3}(a) show that for a monotonically decreasing noise spectrum the CPMG SD method reconstructs the noise spectrum with high precision.
Repeating the procedure for the DYSCO FF reveals the noise spectrum with similar accuracy.
In the next step a Gaussian peak around $\SI{62}{\kHz}$ is added to the initial noise spectrum, which simulates a $^{13}$C Larmor peak originating from surrounding nuclear spins at an external magnetic field of $|B|\approx60$\,\si{Gauss}.
Fig.\,\ref{fig:fig3}(b-d) depicts the reconstruction of the simulated noise spectra with CPMG SD, DYSCO and gDYSCO, respectively. The spectra curves show the reconstruction with (red) and without (green) the addition of experimental uncertainty to the simulated coherence curves.
The gDYSCO scheme results in a superior reconstruction of the $^{13}$C peak, while CPMG SD suffers from a significant systematic bias caused by the higher harmonics and side lobes of the FF.
When statistical noise is added, the limited sensitivity of gDYSCO alters the results, but the peak is still clearly visible as most of the noise affects the monotonous ``background''.
In the CPMG SD case, the noise has a limited effect on this ``background'', but significantly flattens the noise peak. Additionally, it can be seen that the CPMG SD method, in the presence of noise, pushes the peak to slightly higher frequencies.
This is caused by the asymmetry of the filter function and the higher harmonics. %felix x1 (what)

The simulated noise spectrum reconstructed with gDYSCO reproduces the noise peak with much higher precision than DYSCO.
Despite the fact that the main peak of the DYSCO FF is a factor of two narrower than that of the gDYSCO, it still produces a much wider $^{13}$C Larmor peak that is comparable to the one reconstructed by the CPMG SD method.
This indicates a significant contribution originating from the side lobes of the DYSCO FF and highlights the fact that the resolution is not limited by the simple FF FWHM in the case of non-monotonic spectra. Nevertheless, the FWHM gives a bound on the resolution, which is reached when the spectrum does not have a significant peak.
Since both the CPMG FF and the DYSCO FF have feature-comparable side lobes, the resulting spectrum around the $^{13}$C Larmor peak appears similar for both of them, with the CPMG FF higher harmonics contributing as a $\text{2}^{\text{nd}}$ order effect.

\section{Experiment}
\begin{figure}[htbp]
	\begin{center}
		\includegraphics[width=1 \columnwidth]{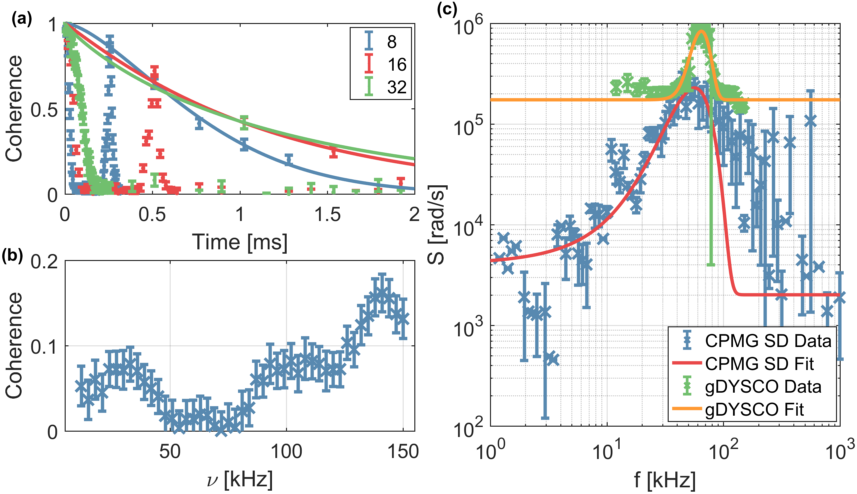}
		\protect\caption{
        (a) Experimental coherence curves for CPMG-N experiments. Short time scales are densely sampled in order to reveal the $^{13}$C collapse. The revival points are also sampled in order to extract the decoherence curve. The envelope functions $\exp\left(-\left(x/T\right)^p\right)$ fitted to the revival points are shown as solid lines.
        (b) Coherence curve obtained from a gDYSCO measurement as a function of the modulation frequency. 
        (c) The reconstructed spectrum using CPMG 2-128 and the spectrum obtained from the gDYSCO measurements. Both spectra were fitted with a Gaussian, the fitted parameters can be seen in Table\,\ref{table:peak}.
        The $^{13}\mathrm{C}$ Larmor peak reconstructed by gDYSCO is narrower than the one obtained by CPMG SD, as predicted from the simulation results (Fig.\,\ref{fig:fig3}(b)). In addition, it can be clearly seen that the limited sensitivity of gDYSCO does not reveal the noise spectrum when its power is below $\approx 10^5\,\frac{\textrm{rad}}{\textrm{s}}$ but results in a higher maximum detected noise spectrum (see appendix for more information).
        }\label{fig:fig4}
	\end{center}
\end{figure}

Experimental confirmation of our findings is obtained by utilizing an ensemble of NV centers coupled to a bath of $^{13}\textrm{C}$ nuclear spins in a ``off the shelf'' diamond sample. The sample is a polished CVD grown single-crystalline diamond sample (Element Six, catalog number 145-500-0248) featuring a natural $^{13}\mathrm{C}$ abundance, a nitrogen concentration of $<1\,\mathrm{ppm}$ and a natural dense NV ensemble with a concentration of $\approx 1 \,\mathrm{ppb}$.
We performed CPMG experiments on two times scales. Short time-scale measurements with high temporal resolution allow to extract the coherence revival caused by the Larmor precession, and thereby the magnetic field. In contrast, long time scale measurements sampling only the Larmor revivals extract the coherence curve envelope.
These two experiment classes allow reconstructing both the $^{13}$C Larmor peak as well as the long-time, low-frequency component of the noise.
Additionally, DYSCO and gDYSCO experiments have been performed, each with a total experiment time of $\tDYSCO = \SI{200}{\micro\second}$.
The experiments were performed by applying the pulse sequences shown in Fig.\,\ref{fig:fig1}. The raw experimental data for both the CPMG and gDYSCO are presented in Fig.\,\ref{fig:fig4}(a,b) and the resulting/reconstructed spectra are presented in Fig.\,\ref{fig:fig4}(c) (data for the normal DYSCO sequence can be found in the appendix).
In order to extract the central frequency and width, the $^{13}$C Larmor peak is fitted by a Gaussian function (Fig.\,\ref{fig:fig4}), (exact values are shown in Table\,\ref{table:peak}). 
For comparison, based on the external magnetic field, the Larmor frequency was determined to be $f_{\text{Larmor}} =  \SI{62.5\,(6)}{\kHz}$.

\begin{table}
	\centering
	\caption{ $^{13}$C Larmor frequency as extracted from the reconstructed experimental noise spectra by means of a Gaussian fit. The width of the Larmor feature is the fitted Gaussian 1$\sigma$}
\begin{tabular}{|c|c|c|}
	\hline
	 &  CPMG  &  gDYSCO \\
	\hline
    f [kHz] & $60.2 \pm 4.2$ & $64.3 \pm 0.5$ \\
	\hline
    Width [kHz] & $29.1 \pm 2.2$ & $13.8 \pm 0.8$ \\
    \hline
\end{tabular}
\label{table:peak}
\end{table}

Fig.\,\ref{fig:fig4}(c) highlights the benefits of the different methods.
Large bandwidth information is recovered by the CPMG SD, spanning a frequency range three orders of magnitude larger than for gDYSCO.
Within this range, the sensitivity dynamic ratio is higher by at least two orders of magnitude, thereby giving a significantly improved estimate on the upper bound of the noise contribution.
However, the ability to resolve narrow resonant features within the spectrum is significantly improved for gDYSCO.

\section{Discussion}
We briefly compare our work to three other recent papers that we have become aware of during the preparation of this manuscript \cite{AXY8,frey_application_2017,norris_optimally_2018}.
In \cite{AXY8} adaptive XY-n sequence (AXY-n) are introduced, which are non-equally-spaced XY-n sequences. The AXY-n sequence also suppresses the higher harmonics present in the CPMG sequence and could, theoretically, be used for noise spectroscopy. However, there are two caveats for using AXY-n compared to gDYSCO: (1) Due to the finite sequence duration, the AXY-n will still have the same side-lobes as CPMG and DYSCO, which will result in decreased spectral resolution; (2) The AXY-n sequence still has other spurious peaks, which might have to be decomposed in a similar SD manner. However, it is possible that the location of these spurious peaks can be controlled well enough for this to be ignored. In addition, DYSCO is composed of continuous driving with frequent phase changes, which increases its susceptibility to pulse errors. Overall, it seems that an AXY SD might have an advantage compared to CPMG SD, but an in-depth comparison encompassing also gDYSCO merits further study.
In \cite{frey_application_2017,norris_optimally_2018}, the authors suggest using discrete prolate spheroidal sequences (DPSS) \cite{DPSS} in order to probe and identify fine spectral features (resonant peaks). The DPSS, like DYSCO and gDYSCO, do not have the higher harmonics (sometimes referred to in the literature as ``spectral leakage'' or ``Gibbs artifacts'') present in CPMG SD. In addition, they have reduced side-lobes, less than CPMG and DYSCO, but more than gDYSCO. In these papers, the authors used simulations in order to characterize the noise spectroscopy using DPSS. Therefore, a direct comparison to gDYSCO in terms of sensitivity, bandwidth and resolution is non-trivial and beyond the scope of the current manuscript.

In conclusion, we have analysed and compared different quantum noise spectroscopy methods for the case of non-monotonic spectra, specifically containing a large and narrow resonant feature.
We found that the gDYSCO scheme provides higher resolution and allows for a more precise detection of distinct features, while spectral decomposition based on CPMG exhibits higher sensitivity and larger bandwidth.
Therefore, this work suggests that the best approach is a combination of different approaches in order to reveal full spectral information of non-trivial noise baths (our results indicate that optimized conditions might increase the sensitivity of gDYSCO, as described in the appendix). This insight could provide an important tool for the study and characterization of a wide range of quantum systems, such as various solid-state defects, super-conducting circuits, quantum dots and trapped ions, leading to a deeper understanding of the relevant physical processes, as well as optimized control schemes for quantum applications such as sensing and quantum information processing. In particular, this method could be used in NMR/MRI measurements, traditionally done with pulsed XY8 sequences, to improve the sensitivity, bandwidth and accuracy. Other works that have found non-monotonous magnetic noise, such as pulsed modulation measurements done with ions\cite{Kotler2011} could also benefit from these new methods.

\section{Acknowledgments}
We would like to thank Stefan Hell for the use of lab equipment and for his support in the project. This work has been supported in part by the Niedersachsen-Israel cooperation program (Volkswagen Stiftung), the Minerva ARCHES Award, the EU (ERC StG), the CIFAR-Azrieli Global Scholars program, the Ministry of Science and Technology, Israel, and the Israel Science Foundation (Grant No. 750/14). Y.R. is grateful for the support from the Kaye Einstein Scholarship and from the CAMBR fellowship.

\section{Appendix}
\subsection{Spectral Decomposition (SD) method for CPMG}
The CPMG-N FF is given by the following formula \cite{dassarma_dd}:

\begin{center}
\begin{tabular}{l|c}
Even N \,& \,$\frac{16}{\pi t \omega^2} \text{sin}^4\left(\frac{\omega t}{4N} \right) \text{sin}^2\left(\frac{\omega t}{2} \right) \,/\, \text{cos}^2\left(\frac{\omega t}{2N} \right)$ \\
Odd N \,& \, $\frac{16}{\pi t \omega^2} \text{sin}^4\left(\frac{\omega t}{4N} \right) \text{cos}^2\left(\frac{\omega t}{2} \right) \,/\, \text{cos}^2\left(\frac{\omega t}{2N} \right)$ \\
\end{tabular}
\end{center}
Numerical study of the functions shows that they have a maximum at approximately $\omega_0 \approx \frac{\pi N}{t}$ with $t$ being the experiment time. This approximation is good for large $N$s; the table below shows the deviation as a function of N:
\begin{center}
\begin{tabular}{r|c}
N\, & $\omega_0 / \frac{\pi N}{t}$ \\
\hline
1\, & 1.48 \\
2\, & 2.30 \\
3\, & 3.21 \\
4\, & 4.17 \\
8\, & 8.09 \\
16\, & 16.04 \\
\end{tabular}
\end{center}
The FWHM of the peak is given by $\Delta \omega \approx \frac{1.77 \omega_0}{N}$.
The area beneath this peak contain approximately 75\% of the total FF area, with about 8\% more in the side lobes, 10\% in the 2$^\text{nd}$ harmonic which is at approximately $\omega_1 \approx 3\omega_0$ and the rest in the higher harmonics.

In order to perform the spectral decomposition analysis, the following prerequisites need to be fulfilled:
\begin{enumerate}
	\item 	There is some cutoff at higher frequencies, and we are probing close enough to it (or after it) with several data points.
For these data points, all of the contributions to $\chi$ come from the main lobe.
	\item	The shortest CPMG data point is taken before any decoherence process affected the NV, so that the coherence is maximal. This is important for the rescaling of coherence curve and therefore $\chi$ and $S$. Of course, a small deviation is not critical.
	\item	The FF main lobe width is small compared to the changes in the spectrum such that the spectrum is approximately linear inside the main lobe. This is much easier to achieve for high CPMG N’s.
    \item 	Alternatively, the spectrum is probed densely, such that adjacent data points have overlapping FF and the spectrum does not change much between them.
\end{enumerate}

The SD is done in two steps: the first step assumes that the main contribution comes from the 1$^\text{st}$ harmonic, which is treated as a rectangular function with the same width and area as the main peak FWHM. An initial spectrum is calculated by approximating Eq.\,(\ref{eq:chi}):
\begin{align}
\chi(t) & \approx \frac{t}{2} \int_{\omega_0-\Delta}^{\omega_0+\Delta} S(\omega) \mathrm{FF}(\omega t) \mathrm{d}\omega \nonumber\\
        & \approx \frac{t}{2} S(\omega_0) \int_{\omega_0-\Delta}^{\omega_0+\Delta} \mathrm{FF}(\omega t) \mathrm{d}\omega \nonumber\\
        & \approx \frac{t}{2} S(\omega_0) \Sigma
\label{eq:approx_chi}
\end{align}
This leads to the 1$^{\text{st}}$ order approximation:
\begin{equation}
S_0(\omega_0) = -\frac{2\chi(t)}{t \Sigma}
\label{eq:initial}
\end{equation}
Where $\Sigma$, the gain, is the area under the main peak and is calculated from $\Sigma = \int_{\omega_0-\Delta}^{\omega_0+\Delta} \mathrm{FF}(\omega t) \mathrm{d}\omega$. 
The second step corrects for the higher harmonics by subtracting their contribution, the 2$^\text{nd}$ order correction is:
\begin{equation}
S(\omega_0) = S_0(\omega_0) - \int_{\omega_0+\Delta}^{\infty} S(\omega) \mathrm{FF}(\omega t) \mathrm{d}\omega
\label{eq:correct}
\end{equation}
This equation is applied iteratively for all points, starting from the highest frequency point downwards.
Due to the overlap in frequency points resulting from the different CPMG-N curves, the resulting spectrum is very ``dense''. In order to extract a more eye-friendly figure, the data is then binned, with the error-bars corresponding to the within-bin spread.

\subsection{CPMG minimum and maximum sensing frequency}
The main peak of the CPMG FF is approximately at $f_0 \approx \frac{N}{2 t} = \frac{1}{4 \tau}$, where $2 \tau$ is the time between pulses in the CPMG experiment. For a pulsed experiment, we have to maintain the condition $2 \tau \gg \tau_{\pi}$, where $\tau_{\pi}=\frac{1}{2 \fRabi}$ is the pulse duration and $\fRabi$ is the Rabi frequency. From these relations we can extract $f_{\text{max}} \ll \fRabi$.

For the minimum frequency, we need to look again at $f_0 \approx \frac{N}{2 t}$. We need to maximize $t$ while minimizing $N$. In the presence of spin bath noise, the coherence time for CPMG is given by $T_2(N) = T_2(1) N^p$ where $T_2(1)$ is the Hahn-Echo time and $0<p<1$ \cite{sousa2009}. Substituting this into the equation above gives: $f_0 \approx \frac{1}{2 T_2(1)} N^{1-p}$. From this, it is clear that the lowest frequency will be given in an Hahn-Echo experiment ($T_2(1) \equiv T^{\text{Echo}}_2$). In practice, this frequency will be given by the longest time in an Hahn-Echo experiment such that we still have a high enough SNR, which is usually 2-4 times $T^{\text{Echo}}_2$. This means that the lowest possible sensing frequency for CPMG is about $f_{\text{min}} \lesssim \frac{1}{2 T^{\text{Echo}}_2}$.

\subsection{Hahn-Echo and DYSCO coherence curves}
\begin{figure}[htbp]
	\begin{center}
 		\includegraphics[width=1 \columnwidth]{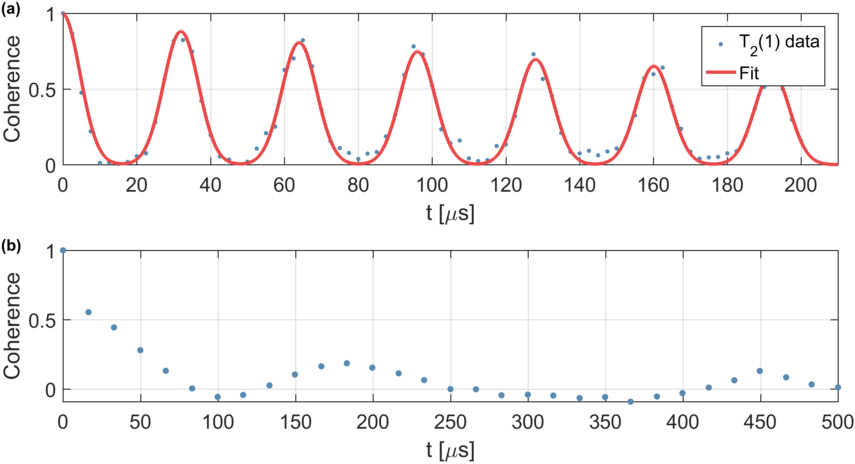}
		\protect\caption{
        (a) The coherence curve obtained for an Hahn-Echo ($T_2(1)$) experiment. The function was fitted to a sum of Gaussians centered around the Larmor revivals multiplied by an envelope: $e^{-(t/T_2)^p} \left(\sum_i{e^\frac{-(t-i*t_l)^2}{2 \sigma^2}}\right)$ where $i$ goes from 0 to 6, and the decay time is $T_2 = \SI{488}{\micro\second}$.
        (b) The coherence curve obtained for a DYSCO experiment with zero sensitivity at all times, as a function of the experiment time. The measurements in the main text were done at $\SI{200}{\micro\second}$, near the first peak in the curve. The occurrence of collapses and revivals are not well understood and will be subject of further investigations.
        %(b) The coherence curve obtained for a Spin-Lock ($T_{1\rho}$) experiment. The function was fitted to $e^{-\frac{t}{T_{1\rho}}}$, and the fitted time is $T_{1\rho} = 7.3\,(9)\,ms$.
        } \label{fig:echo_lock}
	\end{center}
\end{figure}

The lowest frequency that can be probed by the CPMG sequence is $f_{\mathrm{min}}\approx\frac{1}{2 T_2^{\mathrm{Echo}}}$.
We measured $T_2^{\mathrm{Echo}}=488\pm\SI{85}{\micro\second}$ (Fig.\,\ref{fig:echo_lock}(a)) and therefore $f_{\mathrm{min}}\approx\SI{1}{\kHz}$ follows, which is in good agreement with the data presented in Fig.\,\ref{fig:fig4}.
In contrast, the lowest measurable frequency in a DYSCO experiment is equivalent to the inverse of the experiment length, which was $\SI{200}{\micro\second}$.
In order to find the optimal working point we performed a DYSCO experiment with zero sensitivity at all times (Fig.\,\ref{fig:echo_lock}(b)).
The DYSCO experiment length was chosen to be at the first revival point. Working at this point already provided us with enough spectral resolution to perform the experiments and to distinguish the $^{13}$C peak. It is possible to work at the second revival point at $\SI{400}{\micro\second}$ if higher spectral resolution is needed.
Under optimal conditions $t_{\mathrm{DYSCO}}$ approaches $T_{1\rho}$\cite{andrii_DYSCO}, which was measured to be $T_{1\rho} = \SI{7.3\,(9)}{\micro\second}$.
However, the unpolarized nuclear spin in combination with driving field imperfections significantly reduces $t_{\textrm{DYSCO}}$.
A more careful magnetic field alignment, which will significantly increase the nuclear spin polarization, will most likely improve $t_{\textrm{DYSCO}}$ and move it towards $T_{1\rho}$.

\subsection{Noise spectrum simulation vs experiment}
\begin{figure}[htbp]
	\begin{center}
 		\includegraphics[width=1 \columnwidth]{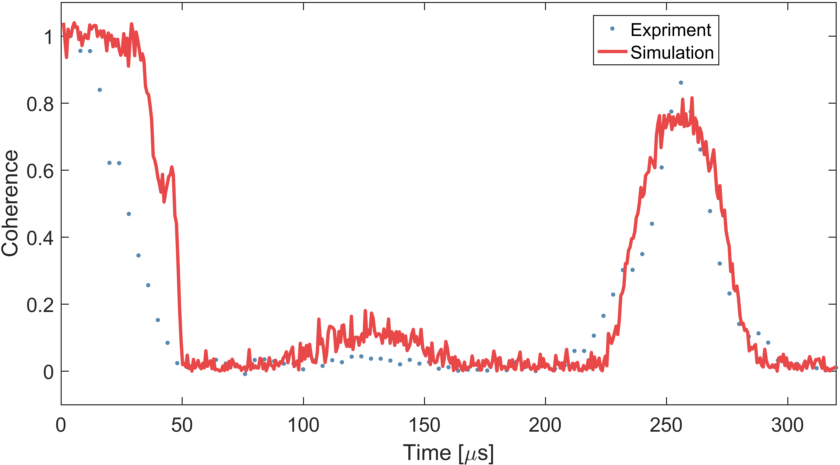}
		\protect\caption{The experimental results for CPMG 8 experiments (in blue) and the simulated CPMG 8 data (in red)} \label{fig:sim_vs_exp}
	\end{center}
\end{figure}
The noise spectrum in the experimental configuration originates from two main sources: the $^{14}$N spin bath which can be modeled as DC centered Lorentzian and the $^{13}$C spin bath which can be modeled as a Gaussian around the Larmor frequency:
\begin{align}\label{sim_noise_spectra}
S(\omega) & = \frac{\Delta_{^{13}\text{C}}^2}{\sqrt{2\pi} \sigma_{^{13}\text{C}}} \text{exp}\left(-\frac{\left(\omega-\omega_{\text{Larmor}}\right)^2}{2 \sigma_{^{13}\text{C}}^2}\right) \nonumber\\
        & + \frac{\Delta_{^{14}\text{N}}^2}{\pi \sigma_{^{14}\text{N}}\left(1+\left(\frac{\omega}{\sigma_{^{14}\text{N}}}\right)^2\right)}
\end{align}
where $\Delta$ are the respective coupling strengths, $\sigma$ are the respective widths and $\omega_{\text{Larmor}}$ is the $^{13}$C Larmor frequency.
In order to obtain a simulated noise spectrum with realistic values, the following procedure was used: a noise spectrum was generated, the CPMG-8 coherence curve was created from the spectrum, and then the simulated CPMG-8 curve was compared to the experimental one. A norm2 minimization problem was then solved by changing the noise parameters in order to minimize the difference between the simulation and the experimental data.
This resulted in the parameters:
\begin{align}
\Delta_{^{13}\text{C}} \approx & \, 500\,\frac{\textrm{krad}}{\textrm{s}}, \;\; \sigma_{^{13}\text{C}} \approx 25\,\frac{\textrm{krad}}{\textrm{s}}, \nonumber\\
\Delta_{^{14}\text{N}} \approx & \, 40\,\frac{\textrm{krad}}{\textrm{s}}, \;\; \sigma_{^{14}\text{N}} \approx 50\,\frac{\textrm{krad}}{\textrm{s}}, \nonumber\\
& \omega_{\text{Larmor}} \approx 392\,\frac{\textrm{krad}}{\textrm{s}}
\end{align}
The comparison between the simulation and experiment can be found in Fig.\,\ref{fig:sim_vs_exp}. While the fine details are different, the main behaviour is similar.
The obtained values are used to generate the spectra shown in Fig.\,\ref{fig:fig3}.

\subsection{Sensitivity and Gain}
\begin{figure}[htbp]
	\begin{center}
 		\includegraphics[width=1 \columnwidth]{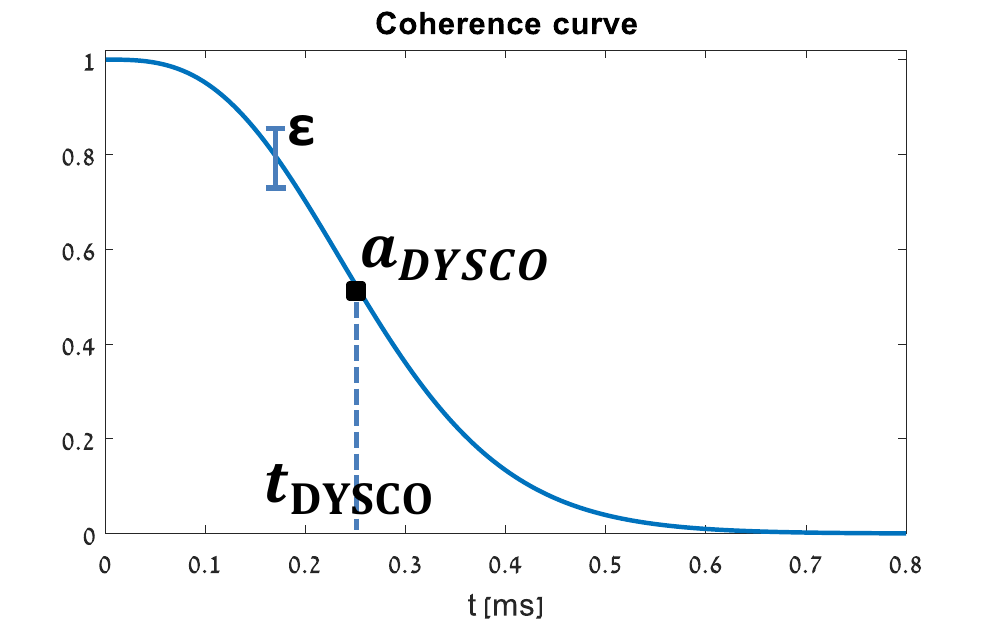}
		\protect\caption{Coherence decay curve. Superimposed are the relevant magnitude of DYSCO maximum contrast, $a_\text{DYSCO}$ and uncertainty, $\epsilon$} \label{fig:contrast_noise_exmple}
	\end{center}
\end{figure}

The sensitivity of the different schemes should take into account the different gains, $\Sigma$, and the measurement uncertainty $\epsilon$ (Fig.\,\ref{fig:contrast_noise_exmple}). The coherence is defined as $C(t)=e^{-\chi(t)}$ and $\chi(t)$ is given in Eq.\,(\ref{eq:chi}). When approximating the FF to Dirac $\delta$ function then $\chi(t)$ is simplified as given in  Eq.\,(\ref{eq:approx_chi}), substituting this into the coherence gives:
\begin{align}
C(t) = e^{\frac{t}{2} S(\omega_0) \Sigma} \nonumber\\
\Rightarrow \text{ln}(C(t)) = -\Sigma \frac{t}{2} S(\omega_0) \nonumber\\
\Rightarrow S(\omega_0) = -\frac{2\text{ln}(C(t))}{t \Sigma} \label{eq:spectrum_gain}
\end{align}
Where $t$ is the experiment duration and $\Sigma$ (gain) is the area underneath the main peak of the FF.
Assuming that the coherence measurement has uncertainty with a standard deviation of $\epsilon$, the coherence that we can detect for CPMG is roughly: $\epsilon \lesssim C(t) \lesssim 1-\epsilon$.
Substitute it into Eq.\,(\ref{eq:spectrum_gain}), the signal range will be:
\begin{equation}\label{min_max_noise_strength_CPMG}
S(\omega_0) = -\frac{2}{t \Sigma_{\text{CPMG}} }
				\left\{
                \begin{array}{ll}
                	\text{ln}(1-\epsilon),\qquad \,\, S_{\text{min}}\\
                  	\text{ln}(\epsilon),\qquad \qquad S_{\text{max}}
                \end{array}
				\right.
\end{equation}
For DYSCO, a specific time, $t_\text{DYSCO}$ is chosen. At this time point, the contrast, $a_\text{DYSCO}$, is smaller than 1 and it is the maximum DYSCO experiment contrast. Repeating the calculation for DYSCO gives:

\begin{equation}\label{min_max_noise_strength_DYSCO}
S(\omega_0) = -\frac{2}{t \Sigma_{\text{DYSCO}}} 
				\left\{
                \begin{array}{ll}
                	\text{ln}(a_\text{DYSCO}-\epsilon),\qquad S_{\text{min}}\\
                  	\text{ln}(\epsilon),\qquad \qquad \qquad \, \, \, S_{\text{max}}
                \end{array}
				\right.
\end{equation}
See Fig.\,\ref{fig:contrast_noise_exmple} for an illustration and Table\,\ref{table:gain} for the values of $\Sigma$.

\begin{table}
	\centering
	\caption{The gain, $\Sigma$, for the different schemes, calculated for the FWHM}
\begin{tabular}{r|c}
Scheme \, & $\Sigma$ \\
\hline
CMPG\, & 0.6 \\
DYSCO\, & 0.35 \\
gDYSCO\, & 0.11 \\
\end{tabular}
\label{table:gain}
\end{table}

If the measurement uncertainty is 3\%, it follows that CPMG can detect two orders of magnitude $\frac{S_{\text{max}}}{S_{\text{min}}} \approx 100$. If a DYSCO measurement is made such that the contrast is dropped to 80\% ($a_\text{DYSCO}=0.8$), then the ratio for DYSCO drops to $\frac{S_{\text{max}}}{S_{\text{min}}} \approx 10$.
Note that the lower gain of gDYSCO allows it to detect a higher noise spectrum, compared to CPMG, by a factor of $\approx 6$, in accordance with the experimental results appearing in Fig.\,\ref{fig:fig4}(c).

\subsection{FFT windows}
In the language of signal processing, gDYSCO is similar to a Gaussian window applied to signal before a discrete fourier transform is applied to it. In a similar manner, the DPSS \cite{frey_application_2017,norris_optimally_2018,DPSS} sequences are similar to a Slepian window \cite{harris_use_1978}. There is an extensive research into the windows which gives the best RMS time-bandwidth product, with the Gaussian window giving an almost optimal value \cite{starosielec_discrete-time_2014}. It is possible to use a better window, such as the Cosine window \cite{starosielec_discrete-time_2014}, to further increase the resolution by a small amount.

\subsection{Confocal setup}
Experiments are performed on a confocal microscope capable of coherent spin manipulation.
Excitation light is provided by a pulsed laser at \SI{532}{\nano\metre} wavelength and a repetition rate of $\SI{30}{\MHz}$.
An acousto-optical modulator (AOM) in a double pass configuration allows for fast on/off switching of the excitation light while guaranteeing an extinction ratio of $\sim 66\,\text{dB}$ in the off state.
The laser is focused onto the diamond using a high NA objective (Leica 1.47). The fluorescence signal in the wavelength window $580-\SI{740}{\nano\metre}$ is collected through the same objective, then directed through a dichroic  mirror and finally detected by means of an avalanche photodiode and stored in a multiple event time digitizer.
The AOM is controlled by an arbitrary waveform generator (AWG, Tektronix), which at the same time is providing the microwave signal to the sample utilizing a microwire antenna.
The fluorescence signal read during the ``readout'' period (Fig.\,\ref{fig:fig1}) is normalized by the counts at the end of the ``initialization'' period which immediately follows. This results in normalization with respect to the excitation laser intensity, greatly reducing the noise caused by fluctuations in the laser intensity, which originate from the laser itself, the AOM and from mechanical instabilities.
Due to the high diamond refractive index, the collection efficiency is $\approx 10\%$\cite{side_collection}. Combined with the coupling to the avalanche photodiode and its quantum efficiency, this resulted in $\approx 2\%$ photon detection efficiency for the fluorescence signal.
The limiting noise source in the setup is shot noise, which is amplified by the low photon detection efficiency\cite{side_collection}. A single readout duration is $\SI{300}{\nano\second}$ with an average of 1 photon per readout. Each experiment is repeated 1 million times, corresponding to a relative shot noise of about 0.1\%. This is to be compared against the contrast, which is on the order of 1-3\%, giving an SNR on the order of 10-30, similar to other works \cite{farfurnik_optimizing_2015}.

\subsection{DYSCO results}
\begin{figure}[htbp]
	\begin{center}
 		\includegraphics[width=1 \columnwidth]{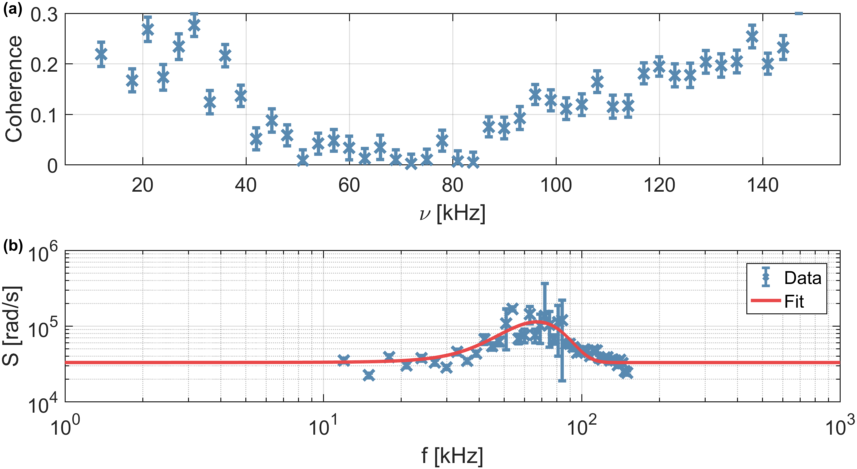}
		\protect\caption{(a) The coherence curve obtained from a DYSCO measurement as a function of the DYSCO modulation frequency. 
        (b) The spectrum obtained from the DYSCO measurements with a fitted Gaussian centered around $\SI{67.1\,(2)}{\kHz}$ with a width of $\SI{17.3\,(2)}{\kHz}$} \label{fig:dysco_spect}
	\end{center}
\end{figure}

The results from the regular DYSCO are presented in Fig.\,\ref{fig:dysco_spect}. These results suffer both from the wide peak due to the side lobes and from the reduced sensitivity of DYSCO.

\bibliography{NV}
\end{document}